\documentclass[pra,twocolumn,superscriptaddress,floatfix]{revtex4}
\usepackage{graphicx,amsfonts,amssymb,amsmath, hyperref}

\newif\ifhyper
\hypertrue
\ifhyper
\hypersetup{
   citecolor = {green},
   colorlinks = {true}, 
   urlcolor = {blue} 
}
\fi

\newcommand{\beq}{\begin{equation}}
\newcommand{\eeq}{\end{equation}}
\newcommand{\beqa}{\begin{eqnarray}}
\newcommand{\eeqa}{\end{eqnarray}}
\newcommand{\ket} [1] {\vert #1 \rangle}

\newcommand{\braket}[2]{\langle #1 | #2 \rangle}

\def\ket#1{\vert#1\rangle}

\def\ipr#1#2{\langle#1\vert#2\rangle}

\def\Longarrow{\protect\@lra}
\def\@lra{\relbar\joinrel\relbar\joinrel\relbar\joinrel%
          \relbar\joinrel\rightarrow}

\begin{document}

\title{Geometric entanglement of one-dimensional systems: \\ bounds and scalings in the thermodynamic limit}

\author{Rom\'an Or\'us}
\affiliation{School of Mathematics and Physics, The University of Queensland, QLD 4072, Australia}
\affiliation{Max-Planck-Institut f\"ur Quantenoptik, Hans-Kopfermann-Stra\ss e 1, 85748 Garching, Germany}

\author{Tzu-Chieh Wei}
\affiliation{Department of Physics and Astronomy, University of British
Columbia, Vancouver, BC V6T 1Z1, Canada}

\begin{abstract}

In this paper the geometric entanglement (GE) of systems in one
spatial dimension  (1D) and in the thermodynamic limit is analyzed
focusing on two aspects. First, we reexamine the calculation of the
GE for translation-invariant matrix product states (MPSs) in the
limit of infinite system size. We obtain a lower bound to the GE
which collapses to an equality under certain sufficient conditions
that are fulfilled by many physical systems, such as  those having
unbroken space (P) or space-time (PT) inversion symmetry. Our analysis justifies the
validity of several derivations carried out in previous works.
Second, we derive scaling laws for the GE per site of infinite-size
1D systems with correlation length $\xi \gg 1$. In the case of MPSs,
we combine this with the theory of finite-entanglement scaling,
allowing to understand the scaling of the GE per site with the MPS
bond dimension at conformally invariant quantum critical points.

\end{abstract}

\maketitle

\section{Introduction} Quantum many-body systems in one spatial
dimension (1D) have proven relevant in physics. For instance,
quantum spin chains have helped to better understand quantum phase
transitions~\cite{qpt, macros}, the renormalization
group~\cite{rg1d}, quantum state transfer~\cite{transfer}, and even
the fundamental limitations of classical and quantum
computers~\cite{complex}. In this respect, the advent of White's
Density Matrix Renormalization Group (DMRG)~\cite{dmrg} was a major
breakthrough since it allowed to simulate many of these systems
efficiently. At its core, DMRG is based on representing
physically-meaningful quantum states in terms of Matrix Product
States (MPSs)~\cite{mps}. Thus, the importance of MPSs can not be
overemphasized: they seem to be the correct low-energy description
of the relevant quantum states of Nature in 1D~\cite{mpsgood}.

Also, in recent years there has been growing interest  in
understanding the properties of entanglement in extended
systems~\cite{entext}. While many studies have focused on bipartite
measures such as the entanglement entropy and single-copy
entanglement~\cite{macros}, there has also been raising interest in
investigating the behavior of  multipartite quantum correlations. In
this respect, among many approaches~\cite{entext}, the so-called
geometric entanglement (GE)~\cite{geometric} has been demonstrated
to be useful in a variety of situations, including 1D systems and
MPSs~\cite{gexy, geometric1d,Shi}, 2D systems~\cite{huang}, and
fully-connected symmetric systems~\cite{geometricother}. However,
and as we shall see here, the study of GE for 1D systems deserves a
more detailed analysis in some cases.

The goal of this paper is to analyze in detail two fundamental
aspects of the GE of 1D systems  in the thermodynamic limit. First,
we reexamine the calculation of the GE of MPSs in the limit of
infinite size, offering a wide perspective of the problem. As a
result, we produce a non-trivial lower bound to the GE. This lower
bound collapses to an equality under certain sufficient conditions
which we shall make precise later. Moreover, we show that some of
these conditions are natural in many physical systems, such as those
having unbroken space inversion or parity (P) symmetry or the larger
spacetime inversion (PT) symmetry~\cite{pt}. Second, we derive
scaling laws for the GE per site of infinite-size 1D systems with
finite but large correlation length. Also, in the case of MPSs we
relate these scaling laws with the theory of finite-entanglement
scaling~\cite{lucapollmann} and obtain scaling laws with the MPS
bond dimension at conformally invariant quantum critical points. In
contrast to the diverging behavior of the entanglement entropy of
with the subsytem size~\cite{macros}, the GE per site is always
bounded, and this gives certain advantage in terms of convergence in
numerics.

The remaining of the paper is organized as follows. In
Sec.~\ref{sec:GE} we review the definition of the geometric measure
of entanglement, or in short geometric entanglement (GE). In
Sec.~\ref{sec:Bounds} we discuss the bounds on GE per site in the
thermodynamic limit. Under certain symmetry, the provided upper
bound becomes the exact expression of GE per site. In
Sec.~\ref{sec:Scaling} we investigate the scaling of GE from three
different perspectives: (a) finite-correlation length, (b) finite
bond-dimension, and (c) finite size. We summarize in
Sec.~\ref{sec:Conclusion}. We relegate certain detailed discussions
on the scaling to Appendix~\ref{app:A} in order to keep the smooth
flow of the main text.
\section{Geometric entanglement in a nutshell} \label{sec:GE}Let us quickly remind
the basics of GE. Imagine that we are given a quantum state
$\ket{\Psi}$ of $N$ parties belonging to a Hilbert space
$\mathcal{H} = \bigotimes_{r=1}^N \mathbb{V}^{[r]}$, where
$\mathbb{V}^{[r]}$ is the Hilbert space of party $r$. This could be,
for instance, the state of a system of quantum spins placed on a
lattice where each party is either a single spin or a block of
spins. Our aim is to focus on the closest normalized product state
of the parties to $\ket{\Psi}$. By ``closest" we mean the normalized
product state $\ket{\Phi} = \ket{\phi^{[1]}}\otimes \ket{\phi^{[2]}}
\otimes \cdots \otimes \ket{\phi^{[N]}}$ that minimizes the squared
distance $|| \ket{\Phi} - \ket{\Psi}||^2$ between $\ket{\Phi}$ and
$\ket{\Psi}$ or, in other words, maximizes the absolute value of
their overlap~\cite{geometric}, $\Lambda_{\max}({\Psi})\equiv
\max_{\Phi}|\ipr{\Phi}{\Psi}|$. Notice that finding the maximizing
product state in this equation is actually akin to a ``mean-field"
approximation to $\ket{\Psi}$.

The larger $\Lambda_{\max}$ is, the less entangled is $\ket{\Psi}$. Thus, the
closest product state approximation to $\ket{\Psi}$ allows us to quantify its
entanglement via the extensive quantity
$E({\Psi})\equiv-\log\Lambda^2_{\max}(\Psi)$~\cite{geometric}, where we have
taken the natural logarithm. As required, $E({\Psi})$ is zero if $\ket{\Psi}$
is unentangled. We can also define the thermodynamic quantity ${\cal E}$ and
its finite-size version ${\cal E}_N$ as
\begin{equation}
{\cal E}\equiv\lim_{N\to\infty}{\cal E}_{N}, \ \ {\cal E}_{N}\equiv
{N}^{-1}E(\Psi).
\label{ene}
\end{equation}
The quantity ${\cal E}$ in the above equation defines the {\it global
geometric entanglement per site}, or {\it density of global geometric
entanglement}. This will be the quantity of interest in this paper.

\section{ Bounds of GE per site in the thermodynamic limit}
\label{sec:Bounds} MPSs offer an accurate description of many
interesting states of 1D quantum many-body systems. For a system of
size $N$ with periodic boundary conditions, these are states defined
as \beq \ket{\Psi} = \sum_{i_1,\ldots,i_N} {\rm tr}( A^{[1] i_1}
\cdots A^{[N] i_N} ) \ket{i_1, \ldots, i_N}, \label{mps} \eeq where
$A^{[m] i_m}$ is a $\chi \times \chi$ matrix at site $m = 1, \ldots,
N$ for each $i_m = 1,\ldots, d_m$, which labels a local basis of the
Hilbert space of dimension $d_m$ at site $m$. Parameter $\chi$ is
called the \emph{bond dimension} of the MPS. It is known that any
spin state can be written as a MPS, albeit the bond dimension may
need to increase with the system size~\cite{complex}. Therefore, MPS
are a good variational family of states to approximate any state,
and in particular, ground states of Hamiltonians with local interactions. If the system is
invariant under translations of one site (TI)~\footnote{The results
of this paper can be easily generalized to other periodicities of
translational invariance.}, then we can assume to have the same
matrices at every site, $A^{[m] i_m} = A^{i_m} \ \forall m$. This
allows us to take the thermodynamic (infinite-size) limit of an MPS
with TI just by considering the same matrices $A^{i_m}$ at all the
infinitely-many sites $m=1,\ldots,\infty$ (see e.g.
Refs.~\cite{idmrgitebd}).

Let us now discuss the calculation of the GE of an MPS in the
thermodynamic limit. One important assumption that we make now is
that the closest product state $\ket{\Phi}$ can be taken to be a
product of identical local states, that is, it fulfills TI:
$\ket{\Phi} = \ket{\phi}^{\otimes \infty}$. This happens to be a
good choice for e.g. the ground states of the transverse-field XY
spin chain, where this ansatz has been verified
numerically~\cite{gexy}. For permutation invariant states, this
ansatz has been proven correct~\cite{geometricother}. However, we
caution that sometimes this assumption does not hold. For instance,
the state $\ket{\Psi} = 2^{-1/2} (\ket{0101...} + \ket{1010...})$ is
manifestly TI, whereas its closest product states (e.g.
$\ket{0101...}$ or $\ket{1010...}$) are not. In such cases a
modified product state ansatz with alternating periodicity (e.g. TI
every 2 sites) is the proper choice, as has been verified
numerically~\cite{geometric1d,Shi, WVG, note}. Moreover, one can
instead consider GE defined with respect to block product states,
and as long as the size of a block is chosen to be a multiple of
possible periods of the possible symmetry-breaking states, such
translation invariant ansatz is expected to hold.

Under the above assumptions, the overlap between an infinite MPS $\ket{\Psi}$
and a product state $\ket{\Phi}$ reads \beq \braket{\Phi}{\Psi} = \lim_{N
\rightarrow \infty} {\rm tr}(B_{\phi}^N), \label{over} \eeq where $B_{\phi}$
is a $\chi \times \chi$ transfer matrix defined as
$B_{\phi}\equiv \sum_i A^i \ipr{\phi}{i}$. Let us call
$\lambda^{\alpha}_{\phi}$ the eigenvalues of matrix $B_{\phi}$ (where $\alpha$
labels the different eigenvalues). Then, we have that ${\rm tr}(B_{\phi}^N) =
\sum_\alpha (\lambda^\alpha_{\phi})^N \sim k(\lambda^1_{\phi})^N$, where the
last approximation is valid in the limit $N\rightarrow \infty$ and where we
assume $\lambda^1_{\phi}$ to be the eigenvalue of largest absolute value and
$k$ its total degeneracy.

The absolute value of the eigenvalue $\lambda^1_{\phi}$ is also known as the
{\it spectral radius} of matrix $B_\phi$, $\rho_{\phi} \equiv
|\lambda^1_{\phi}|$. Another important quantity is the {\it numerical radius}
of $B_\phi$, defined as $w_\phi \equiv \max_{\vec{r}} \left|
~\vec{r}^{\dagger} ~ B_\phi ~ \vec{r} ~\right|$, where $||~ \vec{r} ~|| = 1$.
A crucial property is that the numerical radius of a matrix upper bounds its
spectral radius~\cite{bathia}, \beq \rho_\phi \le w_\phi. \eeq The above
expression turns into an equality if (but not only if) matrix $B_\phi$ is
diagonalizable.

Our bound for the GE is based on considering $w_\phi$ instead of
$\rho_\phi$ as the relevant dominant scale in the thermodynamic
limit for the overlap in Eq.~(\ref{over}). As the latter is not
larger than the former, we have specifically that the maximum
overlap over product states is bounded as \beq
\Lambda_{\max}({\Psi}) \sim \max_{\phi}(\rho_{\phi}^N) \le
\max_{\phi} (w_\phi^N) = \left( \max_{\phi} (w_\phi) \right)^N
\label{si} \eeq where again the first approximation is valid when
$N$ tends to infinity (and up to an irrelevant degeneracy factor
independent of $N$). Therefore, in the thermodynamic limit the
density of geometric entanglement ${\cal E}$ in Eq.~(\ref{ene})
obeys the bound ${\cal E} \ge - 2\log \left( \max_{\phi}
\max_{\vec{r}} \left| ~\vec{r}^{\dagger}~ B_\phi ~ \vec{r} ~\right|
\right)$. The advantage of the latter expression is that it is
variational. Furthermore, the double maximization in the right hand
side of this equation can be further simplified under the assumption
that both maximizations commute~\cite{geometric1d, note2}. In such a
case, the maximization over $\ket{\phi}$ can be done
straightforwardly and we obtain \beq {\cal E} \ge - 2\log \left(
\max_{\vec{r}} | (\vec{r}\otimes\vec{r}^{~*})^\dagger ~ {E} ~
(\vec{r}\otimes\vec{r}^{~*})| \right), \label{gebd} \eeq where $E$
is the zero-dimensional $\chi^2 \times \chi^2$ MPS transfer matrix
$E = \sum_i A^i \otimes (A^i)^*$. The above equation is our main
lower bound on the density of GE.

Let us now discuss the result in Eq.~(\ref{gebd}). To start with,
notice that if the MPS $\ket{\Psi}$ is such that $B_\phi$ is always
diagonalizable $\forall ~ \ket{\phi}$, then the spectral radius
$\rho_\phi$ and the numerical radius $w_\phi$ coincide and
Eq.~(\ref{gebd}) turns into an equality for the density of GE.
Notice also that this is a sufficient condition, but not necessary.
In fact, the ``equality" version of Eq.~(\ref{gebd}) was originally
derived in the second paper of Ref.~\cite{geometric1d},  where
certain assumptions guaranteed the diagonalizability of $B_{\phi}$
for any $\ket{\phi}$. More specifically, it was the (unbroken)
symmetry of the physical system under space inversion (P symmetry)
the key property that guaranteed this diagonalizability. In order to
see this, consider a translation invariant MPS state
 \begin{equation}
 \ket{\Psi} =
\sum_{i_1,\ldots,i_N} {\rm tr}( A^{ i_1} \cdots A^{ i_N} ) \ket{i_1,
\ldots, i_N},
\end{equation}
and its mirror image state (by reversing the ordering in a $N$-spin
basis state without changing the amplitude)
\begin{eqnarray*}
 &&\ket{\Psi'} =
\sum_{i_1,\ldots,i_N} {\rm tr}( A^{ i_1} A^{i_2} \cdots A^{ i_N} )
\ket{i_N,i_{N-1}, \ldots, i_1}\\
&=& \sum_{i_1,\ldots,i_N} {\rm tr}\big( ({A^{ i_N}})^T \cdots
({A^{i_2}})^T({A^{ i_1}})^T\big) \ket{i_N,i_{N-1}, \ldots, i_1}.
\end{eqnarray*}
In order for $|\Psi\rangle=|\Psi'\rangle$, it is sufficient to have
$({A^{s}})^T= U A^s U^{-1}$, for any invertible $U$~\cite{priv,
symmetry}. Thus, $({A^{s}})^T= A^s$ is a sufficient condition for
the space inversion or parity (P) symmetry. Such a condition
guarantees that any $B_\phi$ is automatically symmetric, and
therefore always diagonalizable by some orthogonal transformation. P
(or the larger PT) symmetry of the physical system is thus a
sufficient condition for any matrix $B_\phi$ to be diagonalizable,
which in turn is a sufficient condition for Eq.~(\ref{gebd}) to
collapse to an equality. We stress, though, that a matrix $B_\phi$
may still be diagonalizable even if the MPS does not have P (nor PT)
symmetry.

 The general case may be more intricate.
To see this, notice that any matrix $B_\phi$ (diagonalizable or not)
can be written as $B_\phi = P^{-1}_\phi J_\phi P_\phi$, where
$J_\phi$ is the Jordan normal form of $B_\phi$ \cite{bathia}. Matrix
$J_\phi$ is a direct sum of Jordan blocks, namely $J_\phi =
\bigoplus_\alpha J^\alpha_\phi$ with $(J^\alpha_\phi)_{l,m} =
\lambda^\alpha_\phi \delta_{l,m} + \delta_{l-1,m}$,
$\lambda^{\alpha}_\phi$ being the eigenvalues of $B_\phi$, and $l,m
= 1,\ldots,q_\alpha$ with $q_\alpha$ the size of the Jordan block
$\alpha$. An important theorem in linear algebra states that a
matrix is diagonalizable if and only if $q_\alpha = 1 ~ \forall
\alpha$, that is, all Jordan blocks are trivial. Thus, the
appearance of non-trivial Jordan blocks in the Jordan normal form of
$B_\phi$ makes it non-diagonalizable, which in turn implies  that
Eq.~(\ref{gebd}) may need to remain as a lower bound and not
collapse to an equality, depending on the properties of the Jordan
blocks in $J_\phi$. In such cases, the true value of the GE must be
computed using the left hand side of Eq.~(\ref{si}) and maximizing
the spectral radius $\rho_\phi$ over $\ket{\phi}$, which may be
non-trivial. For instance, consider the case of the
non-diagonalizable
matrix \beq B_\phi = \left( \begin{matrix} \lambda^1_\phi& 0 & 0 \\
0&\lambda^2_\phi & 1 \\ 0 & 0 & \lambda^2_\phi \end{matrix} \right) \eeq which
is already in Jordan normal form and such that its eigenvalues satisfy
$\lambda^1_\phi \ge \lambda^2_\phi$. For simplicity let us assume that these
eigenvalues are real and positive. It is easy to check that the spectral
radius of this matrix is given by $\rho_\phi = \lambda^1_\phi$, whereas the
numerical radius is $w_\phi =  \max(\lambda^1_\phi, \lambda^2_\phi + 1/2)$.
According to this, we have that for $\lambda^1_\phi \ge  \lambda^2_\phi + 1/2$
the numerical and spectral radius coincide and then Eq.~(\ref{gebd}) collapses
to an equality if such a behavior is found for any state $\ket{\phi}$, whereas
for   $\lambda^1_\phi <  \lambda^2_\phi + 1/2$ both radius are different and
therefore Eq.~(\ref{gebd}) must remain as a lower bound.

\section{Scalings of GE per site}\label{sec:Scaling}
In this section we investigate the scaling of GE from three
different perspectives: (a) finite-correlation length, (b) finite
bond-dimension, and (c) finite size.
\subsection{ Finite-$\xi$ scaling of GE per site}
The GE also obeys
precise scaling relations (see e.g. Ref.~\cite{geometric1d}). We now
focus on analyzing the scaling of the density of GE  in
Eq.~(\ref{ene}) as a function of the correlation length $\xi$ for
infinite-size 1D systems, where we assume the correlation length to
be large. Generally speaking, since the GE per site is a density
quantity (similar to the free energy density) the singular part of
GE near criticality is expected to behave as
\begin{equation}
{\mathcal E}(\xi) - {\mathcal E}(\xi=\infty)\sim (b+ b' \log\xi+\cdots)/\xi^d ,
\label{genge}
\end{equation}
where $d$ is the spatial dimension, and $b,b'$ model-dependent coefficients.
In this expression, we have included a first non-trivial logarithmic
correction and assumed that the omitted parts are less singular. In order to
assess the validity of Eq.~(\ref{genge}) we consider some analytical examples,
namely {\it (i)} the case of MPSs for which Eq.~(\ref{gebd}) turns into an
equality, and {\it (ii)} the quantum XY model for spin $1/2$ in the Ising and
XX regimes.

Let us then start with the case of MPSs. Under certain assumptions, it is
possible to find the corrections due to finite correlation length for the GE
per site of an MPS for which the equality version of Eq.~(\ref{gebd}) holds.
First, let us assume that the MPS transfer matrix $E$ obeys an spectral
decomposition $E = \sum_{\alpha=1}^{\chi^2} \mu^{\alpha} \vec{R}^{\alpha}
\vec{L}^{\alpha \dagger}$, where $\vec{R}^{\alpha}$ and $ \vec{L}^{\alpha}$
are respectively the $\alpha$th right and left eigenvectors of $E$ with
eigenvalue $\mu^\alpha$. If the eigenvalues $\mu^\alpha$ are different and
rapidly decaying, then this spectral decomposition can be approximated by $E
\sim \mu^1 (\vec{R}^{1} \vec{L}^{1 \dagger} + (\mu^{2}/\mu^1)  \vec{R}^{2}
\vec{L}^{2 \dagger}) $, and therefore $\mathcal{E} \sim -2 \log |p+
(\mu^{2}/\mu^1) q|$, with $p$ and $q$ some given coefficients. At this point
we recall the standard definition of the correlation length of an MPS, $\xi
\equiv -1/\log|\mu^2 / \mu^1|$. Using this, together with the assumption $\xi
\gg 1$, it is not difficult to arrive at the expression $\mathcal{E}(\xi) \sim
a + b/\xi$, with $a = -2\log |p+q|$ and $b = 2q/|p+q|$. We see thus that
an MPS is in principle capable of handling $O(1/\xi)$ corrections.
Nevertheless, alternative scaling relations may also hold for MPSs if some of
the considered assumptions break down. For instance, it could be the case that
the eigenvalues of the MPS transfer matrix $E$ are not rapidly decaying, and
therefore $\mathcal{E} \sim -2 \log |\sum_{\alpha=1}^{\chi^2} \mu^{\alpha}
q^{\alpha}|$ for the GE (where $q^{\alpha}$ are some coefficients). In
principle, this more general expression could account for all the terms in the
general scaling law from Eq.~(\ref{genge}).

Next, we consider the case of the 1D XY model for spin $1/2$,
defined by the Hamiltonian \beq H_{XY} = -\sum_i \left(\frac{1+r}{2}
\sigma_x^{[i]} \sigma_x^{[i+1]} + \frac{1-r}{2} \sigma_y^{[i]}
\sigma_y^{[i+1]} + h\sigma_z^{[i]} \right), \label{xymodel} \eeq
where $h$ is the magnetic field, $r$ is the anisotropy parameter,
and $\sigma_\alpha^{[i]}$ is the $\alpha$th Pauli matrix at site
$i$. For $0 < r \le 1$ the model belongs to the universality class
of the 1D quantum Ising model (central charge $c=1/2$). In this
regime, in Ref.~\cite{gexy} it was proven that \beq \frac{\partial
\mathcal{E}}{\partial h}\bigg|_{ r \ne 0} \sim -\frac{1}{2 \pi r
\log{2}}\log{|h-1|}, ~~ {\rm for} ~ |h-1| \ll 1. \label{gexy} \eeq
Using the above equation together with the known result $\xi =
1/|h-1|$, we can obtain the dependence of ${\mathcal E}$ on $\xi$.
Integrating the expression in Eq.~(\ref{gexy}) we obtain \beqa
{\mathcal E}(\xi,  r \ne 0) &\sim& a + (b + b' \log \xi)/\xi~~,
\label{IsingXX} \eeqa for $\xi \gg 1$ and where $a, b$ and $b'$ are
some coefficients independent of $\xi$. Notice that this is the same
type of scaling as in Eq.~(\ref{genge}). Also, for $r=0$ the model
belongs to the universality class of the 1D quantum XX model
(central charge $c=1$). In this case, and near $h=1^-$, the GE per
site was shown~\cite{gexy} to behave as ${\cal E}(\xi,r=0)\sim
1/\xi$, where $\xi\sim |1-h|^{-1/2}$. It should be pointed out that
although the XX model near this transition is scale invariant, it is
not conformally invariant. However, near $h=0$ we have seen
numerically that a law like the one in Eq.~(\ref{IsingXX}) seems to
emerge also for the XX model (see Appendix). Remarkably, we see that
all these scaling behaviors accommodate as well in the law hinted in
Eq.~(\ref{genge}).

\begin{figure}
\centerline{\includegraphics[width=9.5cm,angle=0]{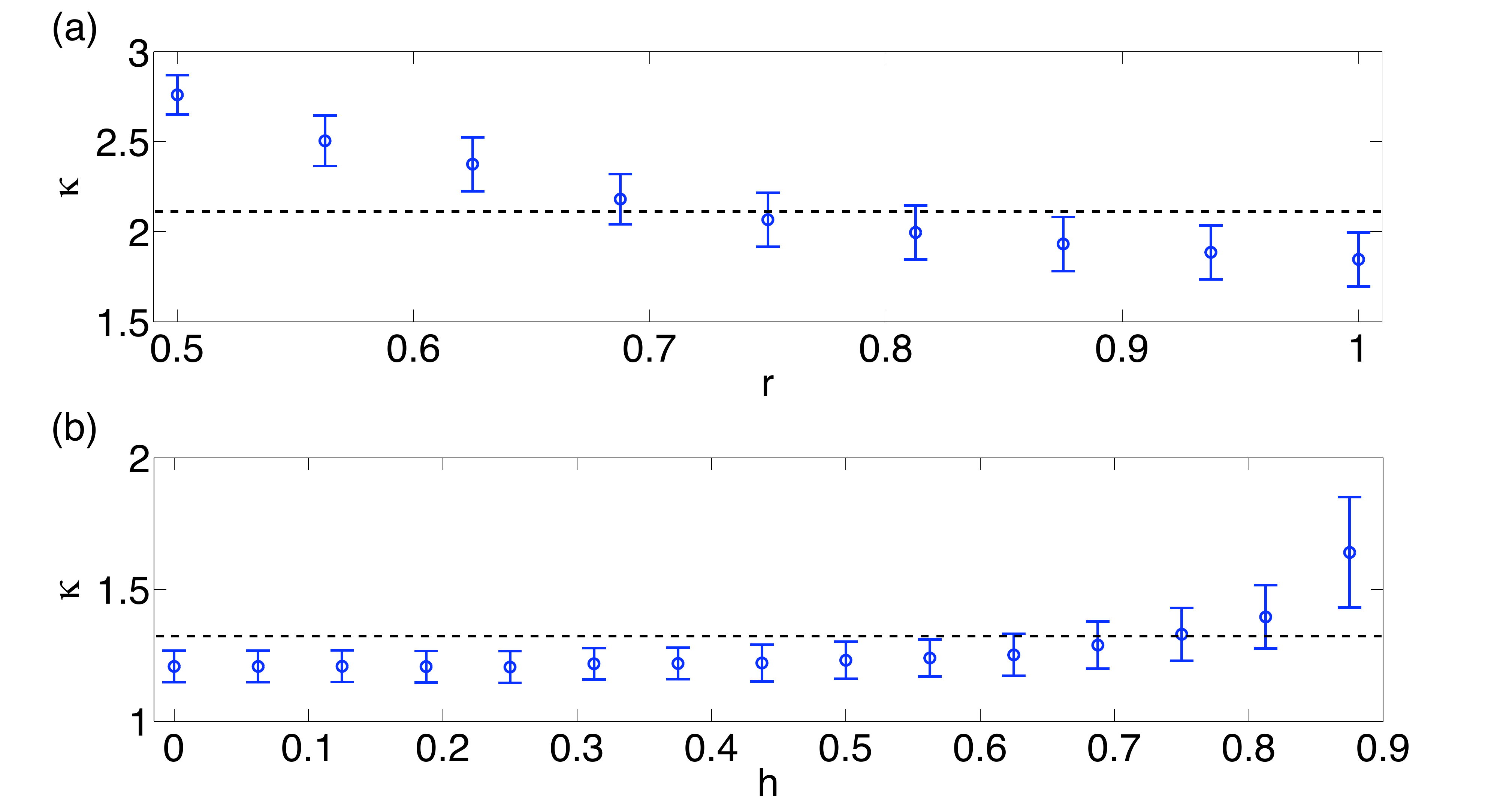}}
\caption{Exponent $\kappa$ as a function of (a) $r$ for the XY model
with $h=1$, and (b) $h$ for the XX model. The values and errorbars
correspond to those in tables \ref{table1} and \ref{table2}. Dotted
lines correspond to the averages over the studied interval, $\kappa
\sim 2.1$ for (a) and $\kappa \sim 1.3$ for (b).} \label{fig:var}
\end{figure}

\subsection{Finite-$\chi$ scaling of GE per site}
An important fact is that the finite-$\xi$ scalings considered here
can be combined with the theory of finite-entanglement
scaling~\cite{lucapollmann}, in turn allowing to understand the
finite-$\chi$ scaling of the GE per site of an MPS with bond
dimension $\chi$. In contrast to the diverging behavior of the
entanglement entropy of with the subsytem size~\cite{macros}, the GE
per site is always bounded above by $\log_2 d$ ebits, where $d$ is
the local Hilbert space dimension, and this gives certain advantage
in terms of convergence in numerics. Knowing that close to a
conformally-invariant quantum critical point the correlation length
obeys the relation $\xi \sim \chi^{\kappa}$ with $\kappa =
6/(c(\sqrt{12/c} + 1))$ ($c$ being the central charge of the
associated universality class)~\cite{lucapollmann}, one obtains the
leading term expression \beq \mathcal{E}(\chi) \sim a + (b + b' \log
\chi )/\chi^{\kappa}~~, \label{fchi} \eeq with $a,b$ and $b'$
coefficients. Remarkably, the validity of this relation is in
agreement with numerical simulations with MPSs for a variety of 1D
spin chains, within some accuracy considerations (see Appendix).
 As illustrated in Fig.~\ref{fig:var}, the
extracted exponent $\kappa$ is very close to the expected value. We
suspect that the deviation is due to the fact that (i) the GE
converges quickly with $\chi$ to the $\chi=\infty$ value and (ii)
the correction is small (as it scales inversely with $\chi$ to some
positive power). The larger deviation for smaller $r$ in
Fig.~\ref{fig:var}a and that at larger $h$ in Fig.~\ref{fig:var}b is
also due to the fact that the $r=0$ and $h=1$ is not a conformally
invariant critical point, and the scaling does not hold. For further
discussions, see the Appendix.

\subsection{Finite-size scaling of GE per site}
 Finally, our scaling laws for finite correlation length
are in accordance with the finite-size scaling behavior found for 1D
systems of size $N$ at the thermodynamic critical point
\cite{geometric1d,Shi,WVG,FrenchGroup}. In that case, it was found
that the GE per site obeyed the law $\mathcal{E}(N) \sim a + b/N$,
with $a$ and $b$ some size-independent coefficients. Up to a
logarithmic factor in some cases, this is the same type of behavior
that we have found here if the role of the correlation length $\xi$
is now played by the size of the system $N$. In general, this is a
manifestation of the well-known property that for a finite system
close to criticality, the size of the system plays the role of an
infrared cut-off in the correlation length.

\section { Conclusions}
\label{sec:Conclusion} Here we have provided a non-trivial lower
bound to the GE of MPSs in the thermodynamic limit. We have
discussed some sufficient conditions under which this lower bound
collapses to an equality, such as space inversion or parity (P)
symmetry, in turn justifying the approach considered in previous
works to compute the GE. We have also derived scaling laws for the
GE per site of infinite-size 1D systems with finite but large
correlation length $\xi \gg 1$. These results have also been related
to scaling of the GE with the bond dimension of MPSs  and the
finite-size scaling of the GE for 1D systems.

\acknowledgements

R. O. acknowledges discussions with J. Fjaerestad, I. P. McCulloch and L. Tagliacozzo, and support from the ARC and UQ. T.-C. W. acknowledges support from NSERC and MITACS.

\appendix
\section{Further discussions on finite-$\chi$ scaling}
\label{app:A}

Here we discuss numerical results that further check the validity of
Eq.~(\ref{fchi}).  We have performed numerical simulations for the spin-1/2
quantum XY model in the regimes corresponding to both the Ising and XX
universality classes (respectively central charges $c=1/2$ and $c=1$), and for the
spin-1/2 antiferromagnetic Heisenberg model ($c=1$). The MPS approximation to the
critical ground state has been obtained using the iTEBD method
\cite{idmrgitebd} for $\chi \in [2,40]$, and the GE has been extracted using
standard optimization tools.

Before presenting our results, a word is in order regarding potential sources of errors. As opposed to other quantities such as the entanglement entropy or the single-copy entanglement, the GE at criticality converges \emph{fast} to a finite value as
$\chi$ (the MPS bond dimension) grows. This is indeed a good feature if we are interested in the GE itself in the limit of infinite bond dimension (as opposed to e.g. the entropy, which tends to be divergent). Our fits try to capture the correction
to this converged finite value. In the vast majority of the cases, such a correction is \emph{small} and
approaches zero as $\chi\rightarrow\infty$ very quickly. Thus, our fits are sensitive to
small numerical errors difficult to control completely. These errors are
accentuated if, on top, the GE itself is very small, as is the case of the XY
model close to $r=0$ and $h=1$ (where the ground state of the system is
separable). If this is the case, then our fits try to capture a tiny correction to a tiny quantity, which may be numerically ill-defined. As we shall see, this seems to be particularly true for the XY model at $h=1$ and as a function of the anisotropy $r$, where the GE itself is around one order of magnitude smaller than for the XX and Heisenberg models. In turn, this also implies that the our most accurate fit to Eq.~(\ref{fchi}) is obtained for the Heiseiberg model, for which the GE is larger. Nevertheless, and in spite of these considerations, our fits succeed in capturing the essential scaling properties of the different systems with good confidence in some regimes.

Keeping the above considerations in mind, our numerical analysis indicates the following:

{\it (i)} For the XY model with $r\neq 0$ and $h=1$ (central charge $c=1/2$ universality
class), the law in Eq.~(\ref{fchi}) fits well the GE for $r$ close to 1,
whereas for $r$ far from 1 the GE decreases quickly towards very small values. This induces to variations in the different parameters of the fit as a function of $r$, which become stronger as the GE approaches zero (see the data in Table \ref{table1}). The observed variations in the parameters are compatible with the laws $a(r) \sim \alpha_a + \beta_a r$, $b(r) \sim -\alpha_b - \beta_b r$, $b'(r) \sim  \alpha_{b'} - \beta_{b'}/r^{1/2}$ and $\kappa(r) \sim \alpha_{\kappa} + \beta_{\kappa}/r^{2}$, for some positive coefficients $\alpha_a, \beta_a, \alpha_b, \beta_b, \alpha_{b'} \beta_{b'}, \alpha_{\kappa}$ and $\beta_{\kappa}$. In spite of this variations, the average behavior of the scaling exponent seems to be $\kappa \sim 2.1$, compatible with the theoretical prediction $\kappa \sim 2$.

\begin{table}
    \begin{tabular}{|c| c |c | c| c| }
    \hline
      $r$ & $a$ & $\kappa$ & $b'$ & $b$ \\
    \hline\hline
~0.5000~ & ~0.015~ & $2.76\pm 0.11$ & $0.010\pm0.001$ & $-0.017\pm 0.002$ \\
~0.5625~ & ~0.017~ & $2.50\pm 0.14$ & $0.012\pm0.001$ & $-0.020\pm 0.002$\\
~0.6250~ & ~0.019~ & $2.37\pm 0.15$ & $0.013\pm0.001$ & $-0.023\pm 0.002$ \\
~0.6875~ & ~0.021~ & $2.18\pm 0.14$ & $0.014\pm0.001$ & $-0.026\pm 0.002$ \\
~0.7500~ & ~0.023~ & $2.07\pm 0.15$ & $0.014\pm0.001$ & $-0.028\pm 0.003$ \\
~0.8125~ & ~0.024~ & $2.00\pm 0.15$ & $0.015\pm0.002$ & $-0.030\pm 0.003$ \\
~0.8750~ & ~0.026~ & $1.93\pm 0.15$ & $0.016\pm0.002$ & $-0.032\pm 0.003$ \\
~0.9375~ & ~0.028~ & $1.89\pm 0.15$ & $0.017\pm0.002$ & $-0.034\pm 0.004$ \\
~1.0000~ & ~0.029~ & $1.85\pm 0.15$ & $0.017\pm0.002$ & $-0.035\pm 0.004$ \\
     \hline
    \end{tabular}
    \caption{Value of the fitted parameters $a$, $b$, $b'$ and the exponent $\kappa$ together with the associated fitting errors, for the XY model with $h=1$, for different values of $r$. The fits are to Eq.~(\ref{fchi}), and in the regime where the GE is not exceedingly small. On average, $\kappa \sim 2.1$. Parameter $a$ has no fitting error since it is extracted directly from our numerical algorithm in the $\chi \rightarrow \infty$ limit.}
    \label{table1}
\end{table}

{\it (ii)} For the XY model with $r = 0$ and $0 < h < 1$ (central charge $c=1$
universality class), we obtain good fits to Eq.~(\ref{fchi}) in the region
close to $h=0$ with values of the scaling exponent around $\kappa \sim 1.2$ (see the data in Table \ref{table2}).
Close to $h=1$, where the ground state becomes very little entangled, the GE
is again too small to extract the scaling behavior with confidence. Furthermore, the
critical point near $h=1$ is not conformally invariant and has dynamical
exponent $z=2$, and hence the relation between $\kappa$ and $c$ (the latter
being not defined in this regime) does not hold. The observed variations in the parameters are compatible with the laws $a(h) \sim \alpha_a - \beta_a h^3$, $b(h) \sim -\alpha_b + \beta_b h^2$, $b'(h) \sim  \alpha_{b'} - \beta_{b'} h$ and $\kappa(h) \sim \alpha_{\kappa} + \beta_{\kappa}\exp(\gamma_{\kappa} h)$, for some positive coefficients $\alpha_a, \beta_a, \alpha_b, \beta_b, \alpha_{b'} \beta_{b'}, \alpha_{\kappa}, \beta_{\kappa}$ and $\gamma_{\kappa}$. Interestingly, we obtain $\beta_{\kappa} \sim 3\times 10^{-5}$ and $\gamma_{\kappa} \sim10$, which involves a smooth behavior of the exponent for low values of $h$, with a rapid increase for large $h$.
Despite of this variations, the average scaling exponent that we obtain in this region is  $\kappa \sim 1.3$, compatible with the theoretical result (which is also $\kappa \sim 1.3$).

\begin{table}
    \begin{tabular}{|c| c |c | c| c|}
    \hline
      $h$ & $a$ & $\kappa$ & $b'$ & $b$ \\
    \hline\hline
~0.0000~ & ~0.155~ & $1.21\pm0.06$ & $0.060\pm0.003$ & $-0.171\pm0.008$ \\
~0.0625~ & ~0.154~ & $1.21\pm0.06$ & $0.060\pm0.003$ & $-0.171\pm0.008$ \\
~0.1250~ & ~0.154~ & $1.21\pm0.06$ & $0.060\pm0.003$ & $-0.170\pm0.008$ \\
~0.1875~ & ~0.152~ & $1.21\pm0.06$ & $0.059\pm0.003$ & $-0.168\pm0.008$ \\
~0.2500~ & ~0.151~ & $1.21\pm0.06$ & $0.059\pm0.003$ & $-0.167\pm0.008$ \\
~0.3125~ & ~0.148~ & $1.22\pm0.06$ & $0.059\pm0.003$ & $-0.166\pm0.008$ \\
~0.3750~ & ~0.145~ & $1.22\pm0.06$ & $0.057\pm0.003$ & $-0.162\pm0.008$ \\
~0.4375~ & ~0.141~ & $1.22\pm0.07$ & $0.057\pm0.003$ & $-0.159\pm0.008$ \\
~0.5000~ & ~0.137~ & $1.23\pm0.07$ & $0.056\pm0.003$ & $-0.156\pm0.009$ \\
~0.5625~ & ~0.132~ & $1.23\pm0.07$ & $0.055\pm0.003$ & $-0.152\pm0.009$ \\
~0.6250~ & ~0.125~ & $1.25\pm0.08$ & $0.054\pm0.004$ & $-0.146\pm0.009$ \\
~0.6875~ & ~0.118~ & $1.29\pm0.09$ & $0.054\pm0.004$ & $-0.142\pm0.010$ \\
~0.7500~ & ~0.108~ & $1.33\pm0.10$ & $0.053\pm0.005$ & $-0.136\pm0.011$ \\
~0.8125~ & ~0.097~ & $1.40\pm0.12$ & $0.051\pm0.006$ & $-0.128\pm0.012$ \\
~0.8750~ & ~0.081~ & $1.64\pm0.21$ & $0.061\pm0.011$ & $-0.133\pm0.021$ \\
    \hline
    \end{tabular}
    \caption{Value of the fitted parameters $a$, $b$, $b'$ and the exponent $\kappa$ together with the associated fitting errors, for the XX model for different values of $h$. The fits are to Eq.~(\ref{fchi}), and in the regime where the GE is not exceedingly small. On average, $\kappa \sim 1.3$. Parameter $a$ has no fitting error since it is extracted directly from our numerical algorithm in the $\chi \rightarrow \infty$ limit.}
    \label{table2}
\end{table}

{\it (iii)} For the spin-1/2 Heisenberg model (central charge $c=1$ universality class), our fit
to Eq.~(\ref{fchi}) seems to indicate that $a \sim 0.259$, $b \sim -0.259 \pm 0.004$, $b' \sim 0.078 \pm 0.010$ and $\kappa \sim 1.25 \pm 0.08$, compatible also with the theoretical result $\kappa \sim 1.3$.

Thus, our data seems compatible with the scaling law in
Eq.~(\ref{fchi}) with values of the scaling exponent $\kappa$ not
too far from the theoretical predictions for conformally-invariant
quantum critical points, within the considerations of numerical
accuracy that we discussed previously. Let us also remark that,
independently of the specific form of scaling, the fact that the GE
converges quickly with $\chi$ to a specific value is also a
remarkable and useful property, as e.g. the infinite-$\chi$ limit
can be easily extracted.


\begin{thebibliography}{000}
\bibitem{note}
Our results can also be generalized to these cases, for example, by considering
the GE per block~\cite{geometric1d}.
\bibitem{note2}
For most cases this turns out to be a correct assumption. However,
it may be possible to find cases where this does not hold.
\end{thebibliography}
\end{document}